%% file: drie.tex
\pdfoutput=1
\documentclass[9pt,twocolumn,twoside]{osajnl}
\usepackage{layouts} 
\usepackage{siunitx}
\usepackage{subfigure} 
\usepackage[colorinlistoftodos,prependcaption,textsize=tiny]{todonotes}
\usepackage{lipsum, tikz}
\usetikzlibrary{shapes,arrows, shadows}

\journal{ao} 

\setboolean{shortarticle}{false} 

\input{Sections/0abstract}

\setboolean{displaycopyright}{false}

\begin{document}

\maketitle
\thispagestyle{fancy}
\ifthenelse{\boolean{shortarticle}}{\abscontent}{}

\input{Sections/1introduction}

\input{Sections/2design}
\input{Sections/3fabmethods}
\input{Sections/4fabrication}
\input{Sections/5fabresults}
\input{Sections/6transmissionmeasurements}
\input{Sections/7twolayer}
\input{Sections/8conclusion}

\section*{Funding Information}
PAG and MDN acknowledge support from NSF CAREER award 1454881. BJK acknowledges support a NASA Space Technology Research Fellowship. NC acknowledges support from a Kavli Institute at Cornell for Nanoscale Science Fellowship.

\section*{Acknowledgments}
This work was performed in part at the Cornell NanoScale Facility, a member of the National Nanotechnology Coordinated Infrastructure (NNCI), which is supported by the National Science Foundation (Grant ECCS-1542081). The authors thank Shawn Henderson for help with the bonding process and Jeffrey McMahon, Stephen Parshley, Jason Glenn, Jordan Wheeler, and Philip Maloney for useful discussions.

\bibliography{Mendeley.bib,sample.bib,Zotero.bib}

\end{document}

%% file: Sections/0abstract.tex
\title{Deep Reactive Ion Etched Anti-Reflection Coatings for Sub-millimeter Silicon Optics}

\author[1,*]{Patricio A. Gallardo}
\author[1]{Brian J. Koopman}
\author[2,3]{Nicholas Cothard}

\author[1]{Sarah Marie M. Bruno}
\author[4,5]{German Cortes-Medellin}
\author[1]{Galen Marchetti}
\author[6]{Kevin H. Miller}
\author[1]{Brenna Mockler}
\author[1,2]{Michael D. Niemack}
\author[4,7]{Gordon  Stacey}
\author[6]{Edward J. Wollack}

\affil[1]{Department of Physics Cornell University 109 Clark Hall, Ithaca, New York 14853, USA}
\affil[2]{Kavli Institute at Cornell for Nanoscale Science, Cornell University, Ithaca, New York 14853, USA}
\affil[3]{Department of Applied and Engineering Physics, Cornell University, Ithaca, New York 14853, USA}
\affil[4]{Department of Astronomy, Cornell University, 220 Space Science Building, Ithaca, New York 14853, USA}
\affil[5]{Department of Electronics and Telecommunications,
School of Engineering, University of Antioquia, Medellin  050010, Colombia}
\affil[6]{NASA Goddard Space Flight Center, Greenbelt, Maryland 20771, USA}
\affil[7]{Cornell Center for Astrophysics and Planetary Science, Cornell University, Ithaca, New York 14853, USA}
\affil[*]{Corresponding author: pag227@cornell.edu}

\dates{Compiled \today}

\ociscodes{(110.6770) Telescopes; (350.1260) Astronomical optics; (350.4010) Microwaves; (040.1240) Arrays.}

\doi{\url{http://dx.doi.org/10.1364/ao.XX.XXXXXX}}

\begin{abstract}
Refractive optical elements are widely used in millimeter and sub-millimeter astronomical telescopes. High resistivity silicon is an excellent material for dielectric lenses given its low loss-tangent, high thermal conductivity and high index of refraction. The high index of refraction of silicon causes a large Fresnel reflectance at the vacuum-silicon interface (up to 30\%), which can be reduced with an anti-reflection (AR) coating. 
In this work we report techniques for efficiently AR coating silicon at sub-millimeter wavelengths using Deep Reactive Ion Etching (DRIE) and bonding the coated silicon to another silicon optic. Silicon wafers of 100 mm diameter (1 mm thick) were coated and bonded using the Silicon Direct Bonding technique at high temperature (1100 C). No glue is used in this process. Optical tests using a Fourier Transform Spectrometer (FTS) show sub-percent reflections for a single-layer DRIE AR coating designed for use at 320 microns on a single wafer. Cryogenic (10 K) measurements of a bonded pair of AR-coated wafers also reached sub-percent reflections. A prototype two-layer DRIE AR coating to reduce reflections and increase bandwidth is presented and plans for extending this approach are discussed.

\end{abstract}

\setboolean{displaycopyright}{false}

%% file: Sections/1introduction.tex
\section{Introduction}
\begin{figure}[ht]
\centering
  \begin{minipage}[t]{\linewidth}
    \includegraphics[width=\linewidth]{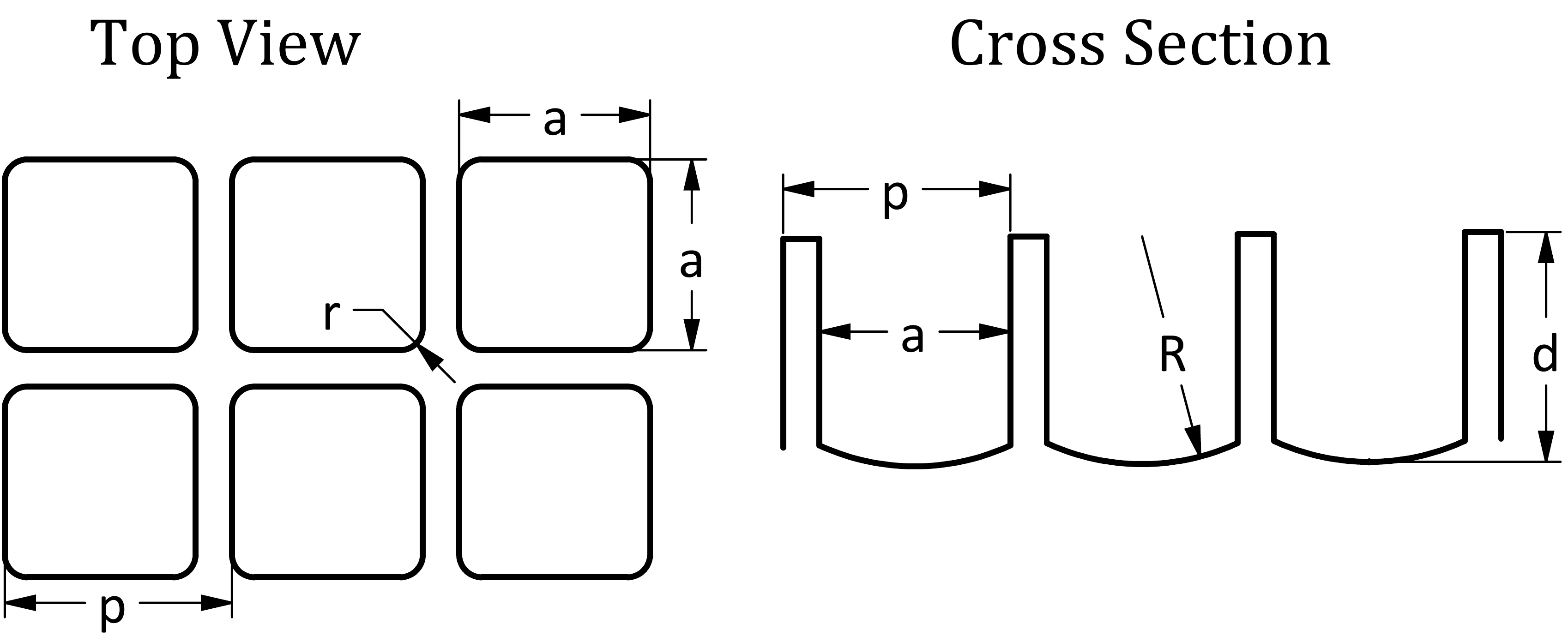}
  \end{minipage}
  \begin{minipage}[b]{\linewidth}
    \includegraphics[width=\linewidth]{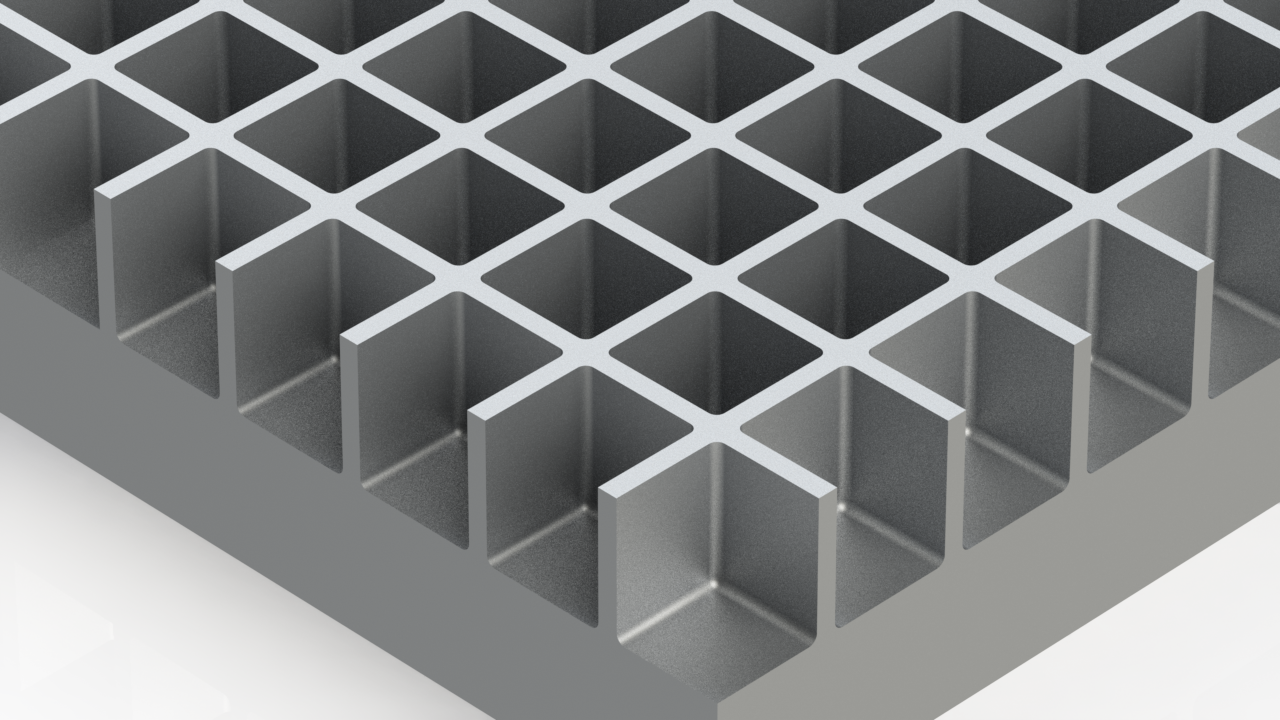}
  \end{minipage}
  \caption{Top: Typical single-layer hole geometry obtained by using DRIE on a polished silicon wafer. Bottom: Schematic view of the simulated dielectric metamaterial structure. Fabrication yields features such as a small radius  of curvature ($r$) on the corners of the squares. The bottoms of the holes are shaped with a finite radius of curvature ($R$). The depth of the hole ($d$) is monitored during the fabrication. The pitch ($p$) and the side of the square of the holes ($a$) are given by the design of the  photolithography mask. Alternative geometries can also be fabricated, see Section \ref{sec:pillars} for an example of  pillars instead of holes.}
  \label{fig:structure}
\end{figure}

Superconducting detector arrays  for millimeter and sub-millimeter astrophysics have become increasingly larger in recent decades (e.g., \cite{henderson_advanced_2016,essinger-hileman_class:_2014,austermann_sptpol:_2012}). This has motivated the need for compact, high throughput optical designs with diffraction limited performance across the entire array. High resistivity silicon is an excellent material for cryogenic dielectric lenses, given its low loss-tangent, high thermal conductivity and high index of refraction \cite{Datta2013Large-apertureWavelengths}. A crucial technology for optical designs with refractive silicon optics is an appropriate anti-reflection (AR) coating. This coating needs to be designed to overcome the $\sim\!$ 30\% Fresnel reflection at each one of the silicon-vacuum interfaces.

An AR layer consists of a layer of a dielectric material that is added on the substrate to reduce the inherent reflection due to the mismatch in the indices of refraction between two media. If the thickness of this dielectric layer is chosen to be $\frac{\lambda}{4}$ in the dielectric media, reflections will create a destructive interference and be impedance matched at freespace wavelength $\lambda = \frac{\lambda_o}{n}$ maximizing the transmission. For the case of silicon ($n_{Si} = 3.4$) and a $\lambda/4$ coating of index of refraction $n$, it can be shown that the reflection coefficient is given by:

\begin{equation}
R = \frac{(n_{Si} - n^2)^2}{(n_{Si} + n^2)^2} \,,
\end{equation}
which is zero for  $n^2 = n_{Si}$ \cite{Hecht2001OpticsEdition, born2000principles}. For silicon, a $\lambda/4$ AR coating gives reflections below $1\%$ over a 25\% bandwidth at normal incidence. Reduced reflections and wider bandwidths can be achieved by the use of multiple coating layers (e.g., \cite{Datta2013Large-apertureWavelengths}). In practice, in order to make an AR coating, a material with the right index of refraction must be found or fabricated to minimize reflections. At mid-IR/optical wavelengths, the AR layer can be applied using a thin film deposition, however at far-IR/millimeter wavelengths the AR layer becomes macroscopic and difficult to apply. Plastic coatings have been explored at millimeter wavelengths  \cite{rosen_epoxy-based_2013,jeong_broadband_2016,lau_millimeter-wave_2006,zhang_new_2009}. Laser milling and laser ablation have been proposed to make simulated dielectric metamaterial AR coatings \cite{drouet_daubigny_laser_2001,matsumura_millimeter-wave_2016,her_microstructuring_1998} but custom high-power optical setups are needed.

Simulated dielectric metamaterial AR coatings consist of sub-wavelength structures on a substrate. This technique allows the designer to tune the effective index of refraction of the AR coating by changing geometric properties of the structure on the substrate. Because metamaterial AR coatings are built from the same material as the substrate, they naturally eliminate differential thermal contraction problems that arise from using materials with different expansion coefficients over large temperature ranges.

Simulated dielectric metamaterial AR coatings have been deployed in Cosmic Microwave Background experiments operating at 90, 150, and 220 GHz \cite{Datta2013Large-apertureWavelengths,henderson_advanced_2016}. In the case of ACTPol, the metamaterial is fabricated using a silicon dicing saw. This method is time-consuming and at shorter wavelengths ($\sim \SI{300}{\micro \metre}$) becomes increasingly challenging to implement since thinner blades are required. For shorter wavelengths the use of Deep Reactive Ion Etching (DRIE) has been studied \cite{wagner-gentner_low_2006,wheeler_antireflection_2014}. DRIE is an anisotropic etching technique used in semiconductor technologies which allows etching of vertical structures of scales ranging from hundreds of nanometers  to hundreds of microns in size.

Since DRIE is typically applied on a flat silicon wafer, we explore the  Silicon Direct Bonding (also called fusion bonding \cite{Gosele1998SemiconductorBonding}) technique, where two polished silicon pieces are bonded together by using a chemical treatment and high temperature  ($1100\ \mathrm{C}$) annealing. Here, the bonding is done via covalent bonds in the silicon-native oxide interface. No glue is used in this process.

In this work we report the manufacture of a 100 mm diameter (\SI{500}{\micro \metre} thick) flat wafer coated on both sides with a square grid structure which achieved reflections lower than $1\%$ at $925$ GHz as measured by a Fourier Transform Spectrometer (FTS). We also report the fabrication of a sample composed of two wafers (\SI{1}{\milli \metre} thick) that were coated on one side each and were bonded  using the  Silicon Direct Bonding technique; this sample was measured to have less than $1\%$ reflections at cryogenic temperatures. Finally, we report on a prototype two-layer DRIE AR coating.

%% file: Sections/2design.tex
\begin{figure}
\centering
    \includegraphics[width=\linewidth]{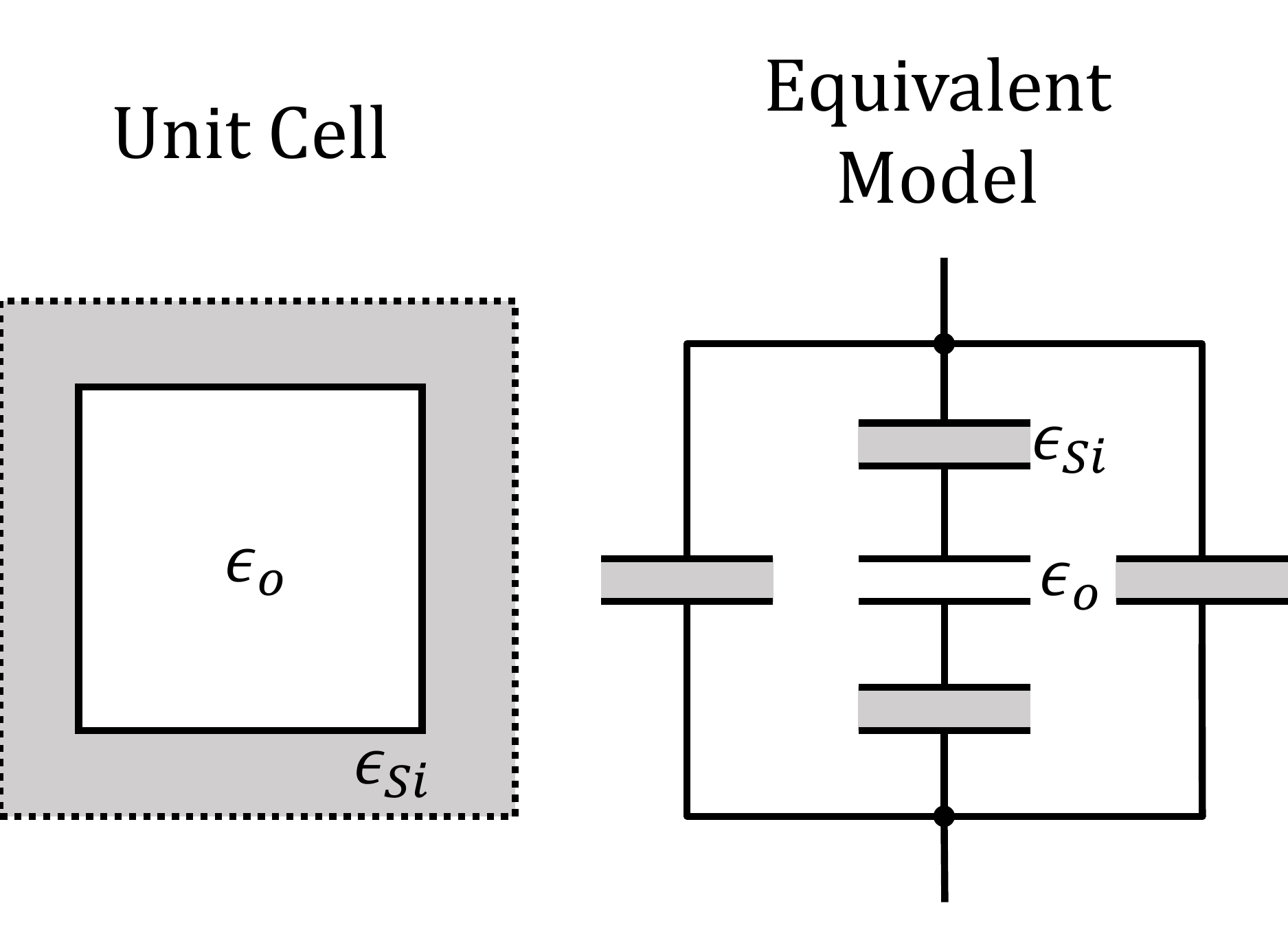}
  \caption{Left: Top view of a unit cell of the metamaterial structure. Right: Equivalent capacitive circuit  \cite{biber_design_2003} used to extract the effective dielectric constant.}
  \label{fig:approximation}
\end{figure}

\section{Design}
The single-layer metamaterial design consists of a grid of sub-wavelength square holes as shown schematically in Figure \ref{fig:structure}. It can be shown that the symmetry of such a grid has zero cross polarization at normal incidence \cite{mackay_proof_1989-1}.  For a pitch much smaller than the wavelength, the structure presents an effective dielectric constant, $\epsilon_{eff}$, to incoming radiation \cite{choy_effective_2015}.

In the early design stages, an analytic Effective Medium Theory \cite{rytov1956electromagnetic} approximation is used to determine the effective dielectric constant of the metamaterial structure. In particular, the capacitive model proposed in \cite{biber_design_2003}  is used to model the square structure depicted in Figure \ref{fig:structure} for $R=\infty$ and $r = 0$. Here the slabs of material in the square structure are treated as the dielectric inside a grid of capacitors as shown in Figure \ref{fig:approximation}. The effective dielectric constant of the structure is approximated  according to \begin{equation}
\epsilon_{eff}=\epsilon_{Si}(1-a/p) + \frac{\epsilon_{Si}\epsilon_o a/p}{\epsilon_{Si} a/p + \epsilon_o(1-a/p)}
\label{eq:eeff}
\end{equation}
where $\epsilon_{Si}$, and $\epsilon_o$ are the dielectric constants of silicon and empty space, $a$ and $p$ are the size of the etched square and the pitch of the structure as defined in Figure \ref{fig:structure}.

For silicon, Equation \ref{eq:eeff} reaches the optimum $\epsilon_{eff} = \sqrt{\epsilon_{Si}}$ when $a/p \approx 0.79$. This ratio was used as a starting point for  electromagnetic simulations using CST Microwave Studio \cite{_computer_????}. These simulations take into account geometrical features that appear in practice after the etching is done. Simulations include the effect of the small radius of curvature ($r$) on the vertex of the square hole and the curvature at the bottom of the hole ($R$). Simulations also took into consideration the \SI{1.5}{\micro\metre} difference between pattern sizes in the mask and in the silicon due to the photolithography and etching. Target feature sizes are summarized in Table \ref{tab:metrology}.

%% file: Sections/3fabmethods.tex
\section{Fabrication methods}
Two fabrication techniques that are common in the semiconductor industry were used: Deep Reactive Ion Etch (DRIE) and Silicon Direct Bonding.

\subsection{Deep Reactive Ion Etch}
\begin{figure}[t]
\includegraphics[width = \linewidth]{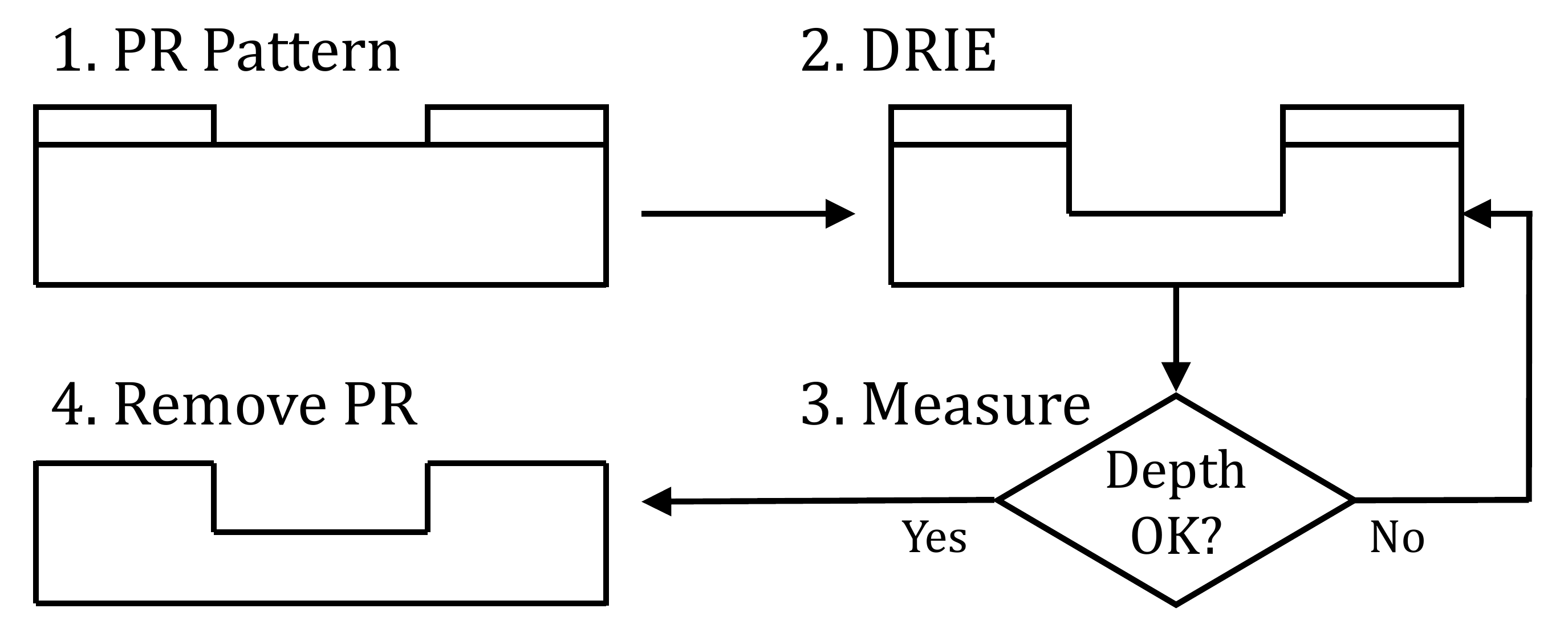}
\caption{Processing steps to define the metamaterial structure on silicon using DRIE: 1. Photoresist is patterned on a flat silicon wafer. 2. DRIE is used to etch areas not covered by the photoresist. 3. An optical profiler is used to measure the achieved depths and decide whether to terminate the process. 4. Photoresist is removed.}
\label{fig:DiagramDRIE}
\end{figure}

Deep Reactive Ion Etch is a process where the kinetic energy from a plasma is used to activate a chemical reaction that carries out the etching. Various chemistries have been proposed, but the Bosch process is the most common \cite{laermer_method_1996}. It requires a mixture of two gasses.  One gas ($SF_6$) enables the  etching and a the second gas ($C_4F_8$) deposits to act as a passivation layer to inhibit sidewall etching. Ion bombardment from the plasma gives the energy to enable the reaction and privileges the vertical direction as walls show a small cross section to the incoming ions. When $SF_6$ is introduced, a very short time is dedicated to remove the passivation layer from the bottom of the hole, the rest of the time is used to etch the silicon and the passivation layer deposited on the walls. In practice, a photoresist (PR) layer is used to create an etch mask, defining the  pattern to be etched. Figure \ref{fig:DiagramDRIE} shows a schematic diagram of the process. See  \cite{Karouta2014AEtching, Wu2010HighReview} for reviews.

\subsection{Silicon Direct Bonding}
Silicon wafer bonding is the process where two mirror polished flat and clean silicon wafers are locally attracted to each other by van der Waals forces and adhere or bond. Typically a heat treatment makes the bonding permanent. This technique is used in areas of silicon-on-insulator (SOI) devices and silicon-based sensors and actuators. In general, silicon wafer bonding involves the following steps \cite{Gosele1998SemiconductorBonding}:

\begin{figure}[t]
\centering

\tikzstyle{block} = [rectangle, draw, fill=black!00, 
    text width=5em, text centered, rounded corners, minimum height=4em, line width = 1pt]
\tikzstyle{line} = [draw, -latex', line width = 1pt]
    
\begin{tikzpicture}[node distance = 3.0cm, auto]

    \node [block] (base) {Base bath $NH_4OH$};
    \node [block, right of=base, node distance = 2.5cm] (rinse1){$H_2O$ \\ Rinse};
    \node [block, right of=rinse1, node distance=2.2cm] (acid) {Acid bath\\ HCl};
    \node [block, right of = acid, node distance = 2.5cm](rinse2){$H_2O$ \\ Rinse};
    \node [block, below of = rinse2, node distance = 2.0cm] (centrifuge) {Centrifuge};
    \node [block, left of=centrifuge, node distance=2.4cm] (contact) {Contact};
    \node [block, left of = contact, node distance = 2.3 cm] (IR){IR inspector};
    \node [block, left of = IR, node distance = 2.3 cm](bake){Furnace process 1200C};
    \path [line] (base) --node [above, text width=0.5cm, align=center] {10 min} (rinse1) ;
    \path[line] (rinse1) -- node {} (acid) ;
    \path[line] (acid) -- node [above, text width=0.5cm, align=center] {10 min} (rinse2);
    \path [line] (rinse2) --node  {} (centrifuge);
    \path[line] (centrifuge) -- (contact);
    \path[line] (contact) -- (IR);
    \path[line] (IR) -- (bake);
\end{tikzpicture}

\caption{Diagram showing steps in the bonding process. First a base bath is followed by an acid bath and rinse. After drying, the wafers are put into contact and inspected with an infra-red camera. The process is reversible at this stage and if the IR inspection shows too many bubbles between the wafers, the process can be repeated. Finally the wafers are pressed and put in the high temperature furnace.}
\label{fig:BondingSchematic}
\end{figure}
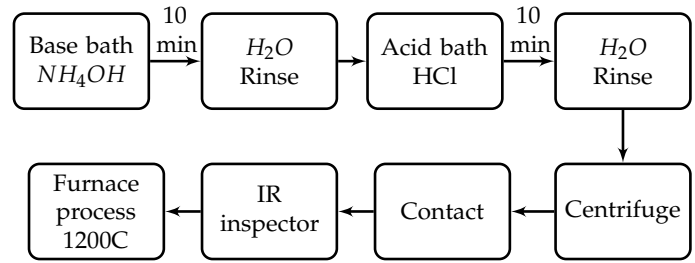

\begin{figure}[!h]
\centering
\includegraphics[width = \linewidth, angle=180]{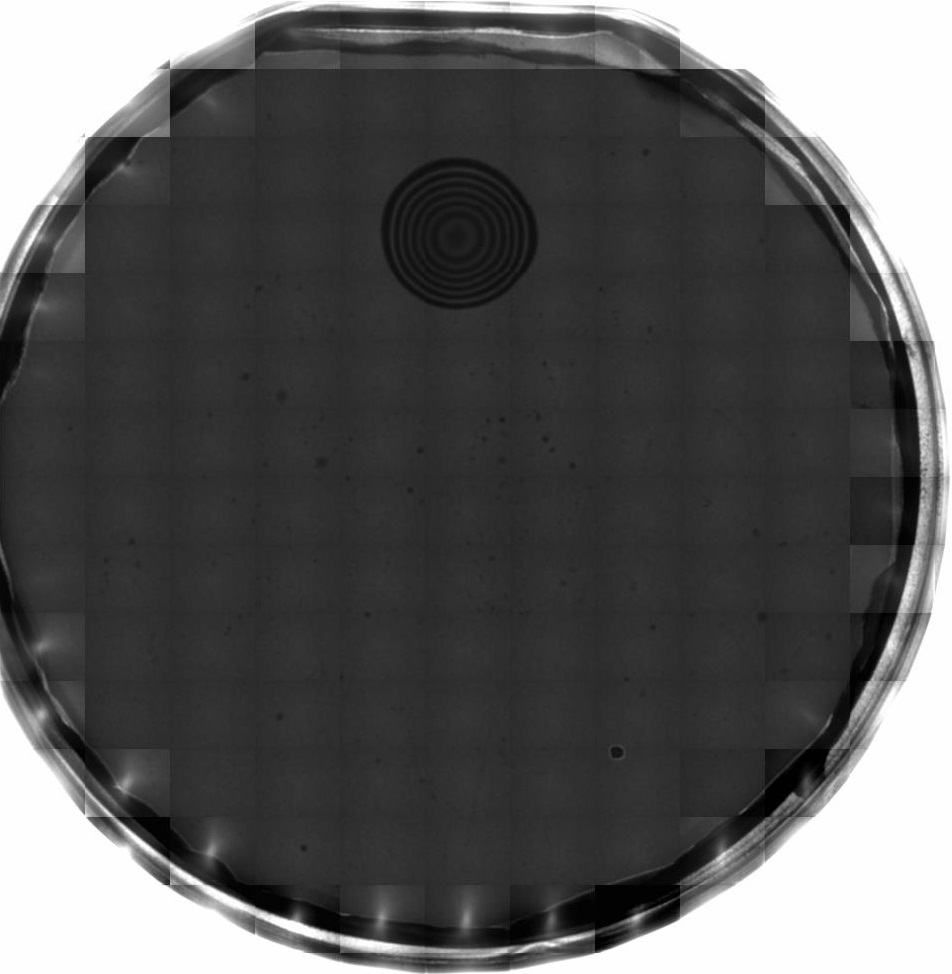}
\caption{IR transmission image of the bonded sample. Imperfections in the bonding are seen as dark spots. In the lower-center region of the image a $\sim\!1cm$ sized bubble is observed. This region didn't achieve complete bonding, probably due to a dust particle. A one micron sized dust particle can result in a bubble of several millimeters in size due to the high stiffness of silicon. The fringes can be used to estimate the bubble thickness, in this case giving an upper limit of $\sim\!4$ microns. The mosaic pattern is caused by the stitching of multiple images in the measuring apparatus.}
\label{fig:IR_Bonded}
\end{figure}

\begin{figure*}
\begin{minipage}{0.5\textwidth}
\includegraphics[width = \textwidth]{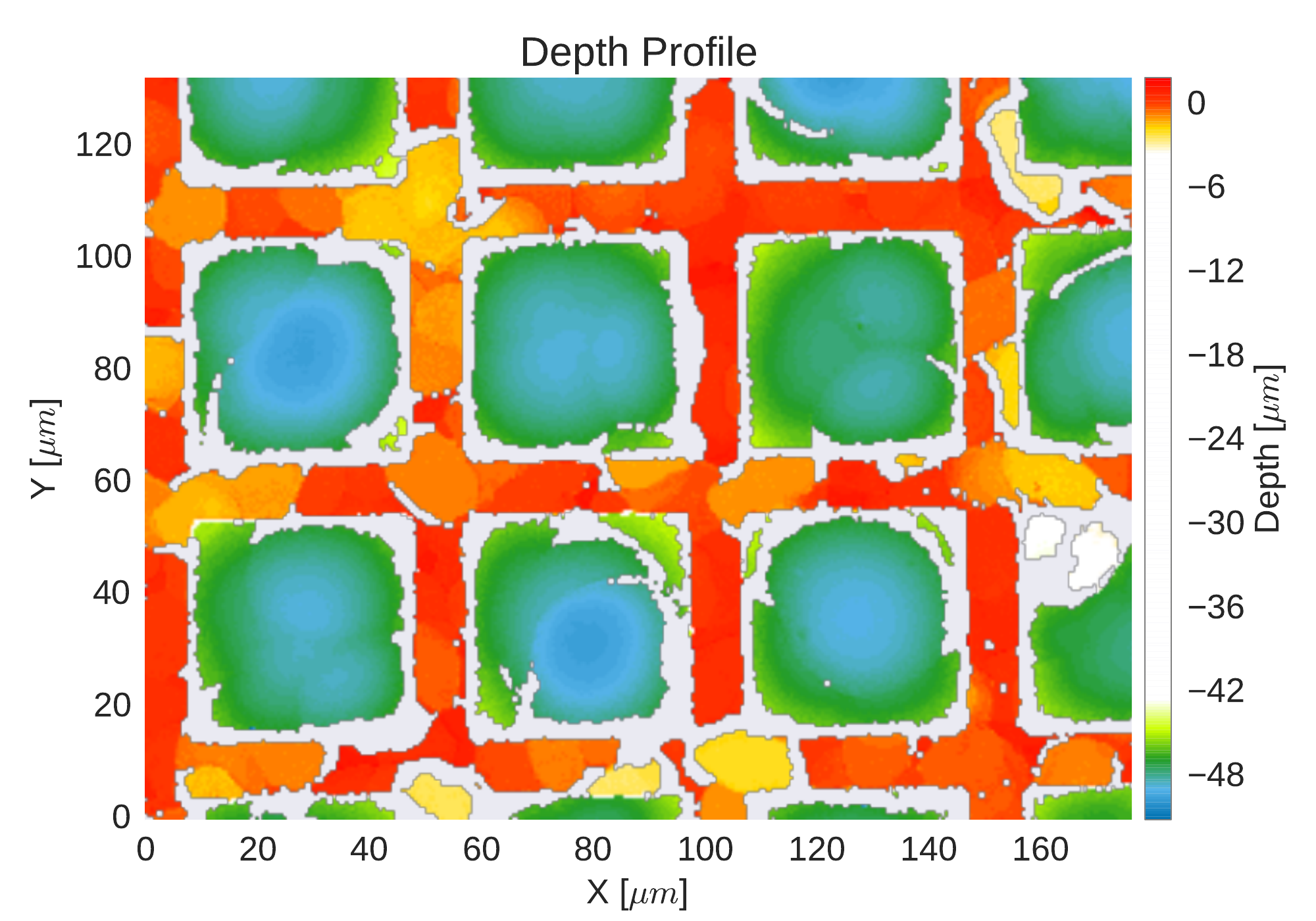}
\end{minipage}
\begin{minipage}{0.5\textwidth}
\includegraphics[width = \textwidth]{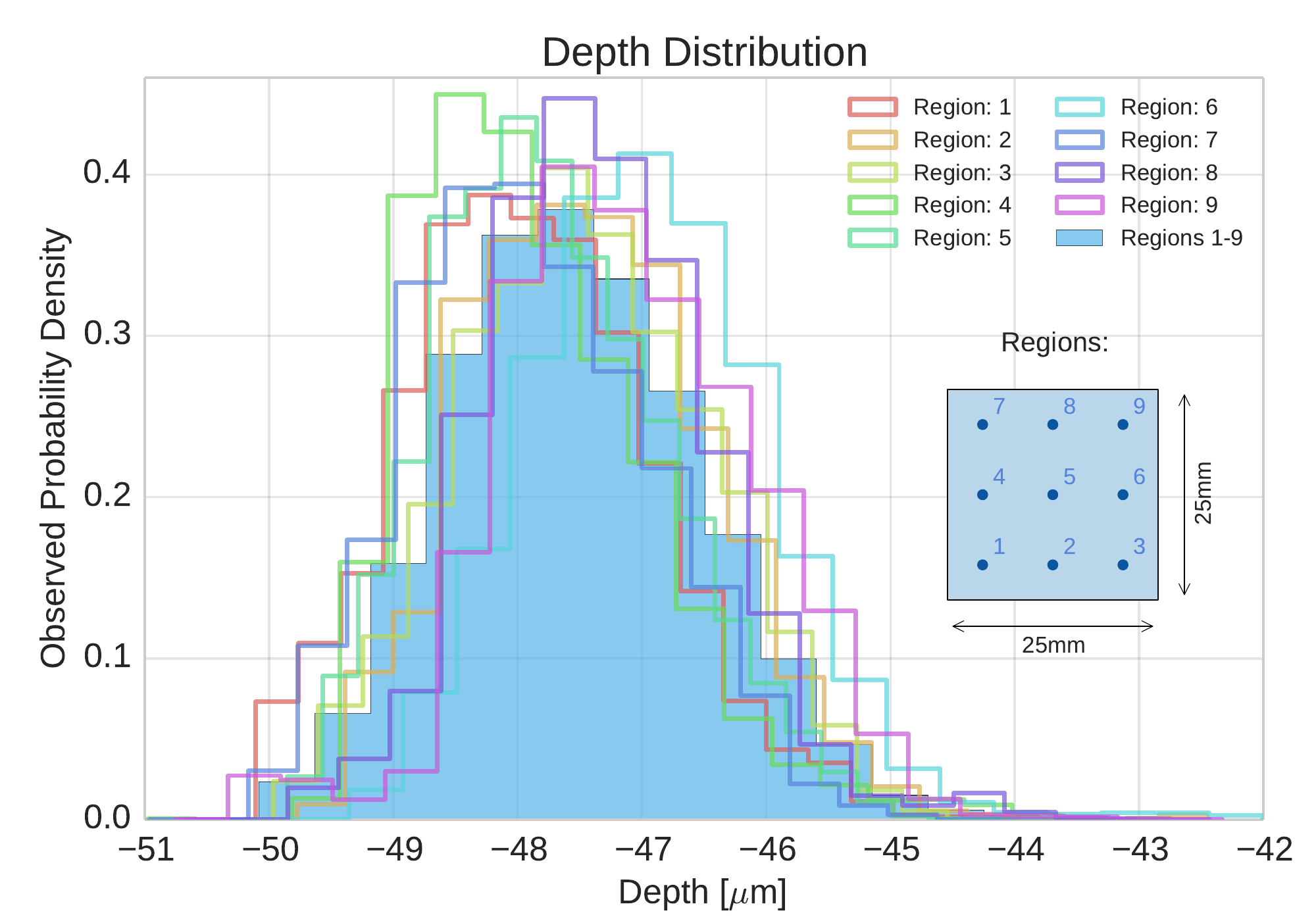}
\end{minipage}
\caption{Etched structure as measured with an optical profiler. Left: One typical profile of a sample that was etched on the unpolished side. Surface roughness was $\sim\!1\mu m$. Surface imperfections expand throughout the etching process leaving different features at the bottom of a trench. This effect is mitigated by using a polished wafer. Right: Histogram of the depths over a 1 square inch silicon piece. Each histogram represents the depth distribution as extracted from the profilometer data in a particular region of a \SI{25}{\milli \metre} silicon piece. Solid histogram represents the aggregate of nine measurements done over this square.}
\label{fig:metrology}
\end{figure*}

\begin{enumerate}
\item The surfaces, which need to be mirror polished are prepared for the bonding process. The interface needs to have some oxide, this can be achieved by thermal oxidation or just a native oxide layer. The wafer interface needs to be coated with monolayers of water which are likely to be adsorbed during the rinse cycles   \cite{christiansen_wafer_2006}.
\item Polished wafers need to be brought into contact at room temperature. A clean environment is critical in this step, as a dust particle of $\sim$ \SI{1}{\micro\metre} in size can lead to a bubble at the bonded interface of $\sim 1\ \mathrm{mm} $ \cite{tong1999semiconductor}. Portable clean room setups (not explored in this work) have been proposed to control this effect \cite{stengl_bubble-free_1988}.
\item After contact, a first stage of (reversible) bonding takes place. Room temperature bonding is given by van der Waals interactions in the silicon interface. A high temperature bath is needed now to make this bonding permanent. Commercial applications are typically treated at  $1100\ \mathrm{C}$.
\end{enumerate}

A variety of chemical mechanisms have been proposed to explain the bonding action at the interface \cite{stengl_model_1989}. A simplified picture of the underlying chemistry of the process can be described with the reaction \cite{Gosele1998SemiconductorBonding}:  
\begin{equation}
2 H_2O + Si \rightarrow SiO_2 + 2H_2
\end{equation}
which occurs at a couple hundred degrees. Higher temperatures allow re-flow of the oxide layer, which helps to fill in the gaps between the silicon wafers. In practice, a chemical treatment is used to remove metal contamination on the wafers and an IR inspector can be used to monitor bubble formation before putting the samples in the high temperature furnace. Figure \ref{fig:BondingSchematic} shows a schematic diagram of the process. See \cite{christiansen_wafer_2006,haisma_direct_2007,tong1999semiconductor,masteika_review_2014}  for reviews on silicon wafer bonding.

%% file: Sections/4fabrication.tex
\section{Fabrication}

Samples were etched after being lithographically patterned. The etching was done using DRIE and an optical profiler was used to monitor the depth of the holes during the process. One 0.5 mm thick silicon wafer was used. Both sides were etched. This sample was used  as a control sample to characterize the performance of the AR structure alone. Two 1 mm thick silicon wafers were etched only on  one side. These two samples were later bonded and transmission of this composite sample was measured.

\subsection{Deep Reactive Ion Etch}
The photolithography mask in use had the same pitch but bigger hole width ($a$) than the target geometry to account for edge effects in the lithography process, since features are expanded during the etching. The mask had  the following geometry: $a = \SI{42.8}{\micro\metre}$, $p = \SI{50.0}{\micro\metre} $. A contact aligner was used to expose the photoresist. After development, etching was performed with the UNAXIS 770 Deep Silicon Etcher available at the Cornell NanoScale Facility (CNF) which gave etch rates of $\sim\! \SI{0.3}{\micro\metre}$ per loop as measured with a Zygo NewView 7300 optical surface profiler (also at CNF). In Figure \ref{fig:DiagramDRIE} a schematic of the process is shown.

\begin{table}[htbp]
\centering
\caption{\bf Design and Measured AR structure parameters.}
\begin{tabular}{ccc}
\hline
Variable  & Measured Value [$\SI{}{\micro\metre}$] & Design Target [$\SI{}{\micro\metre}$]\\
\hline
a & $41.4\pm 0.3$ & 41.3\\
r & $2\pm 0.3$   & 2\\
p & $50\pm 0.3$ & 50\\
R & $68\pm 3$  &$85$\\
$d_{bonded}$ & $47.3 \pm 1$ &48.5\\
$d_{AR}$ & $45.7\pm 1$ & 48.5\\

\hline
\end{tabular}
  \label{tab:metrology}

  Parameter names are defined according Figure \ref{fig:structure}. $d_{bonded}$ and $d_{AR}$ refer to the average hole depth in the two samples that were fabricated in the single wafer and bonded double wafer cases respectively. Differences in $R$ can be approximated by a small change in the effective depth.
\end{table}

\begin{figure*}[!htb]
\begin{minipage}{0.5\textwidth}
\includegraphics[width=\textwidth]{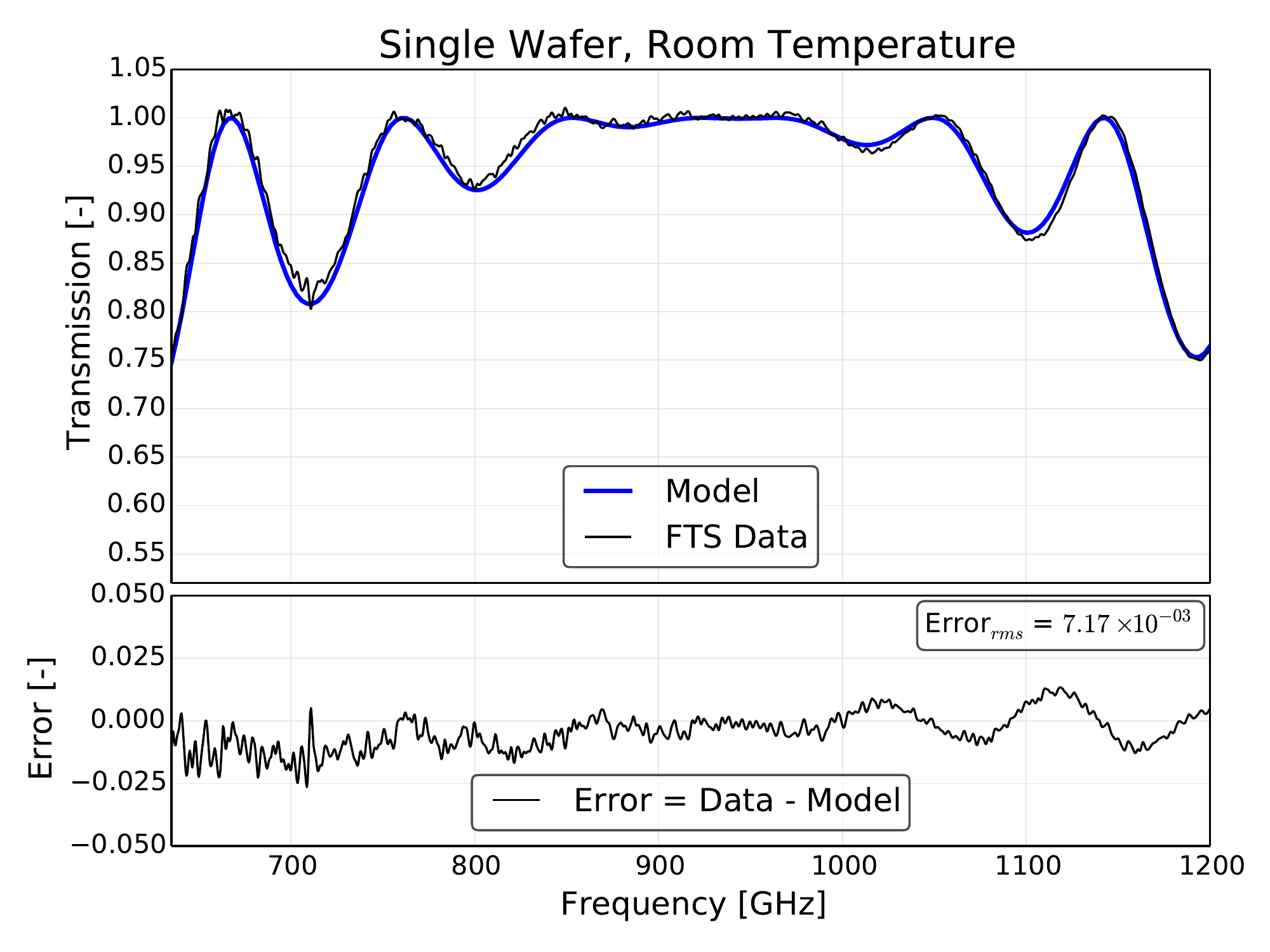}
\end{minipage}
\hspace{\fill}
\begin{minipage}{0.5\textwidth}
\includegraphics[width=\textwidth]{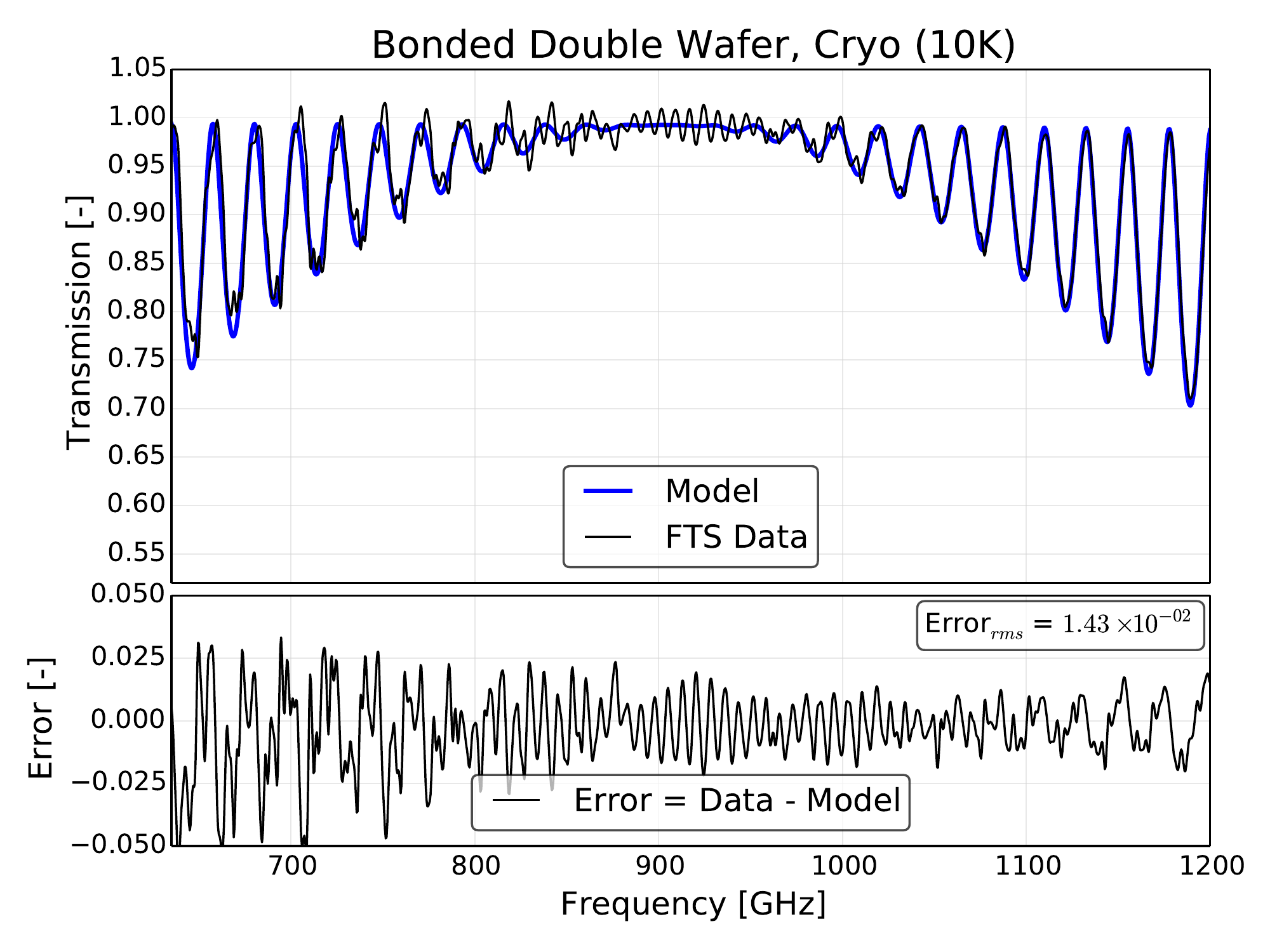}
\end{minipage}
\begin{minipage}{0.5\textwidth}
\includegraphics[width=\textwidth]{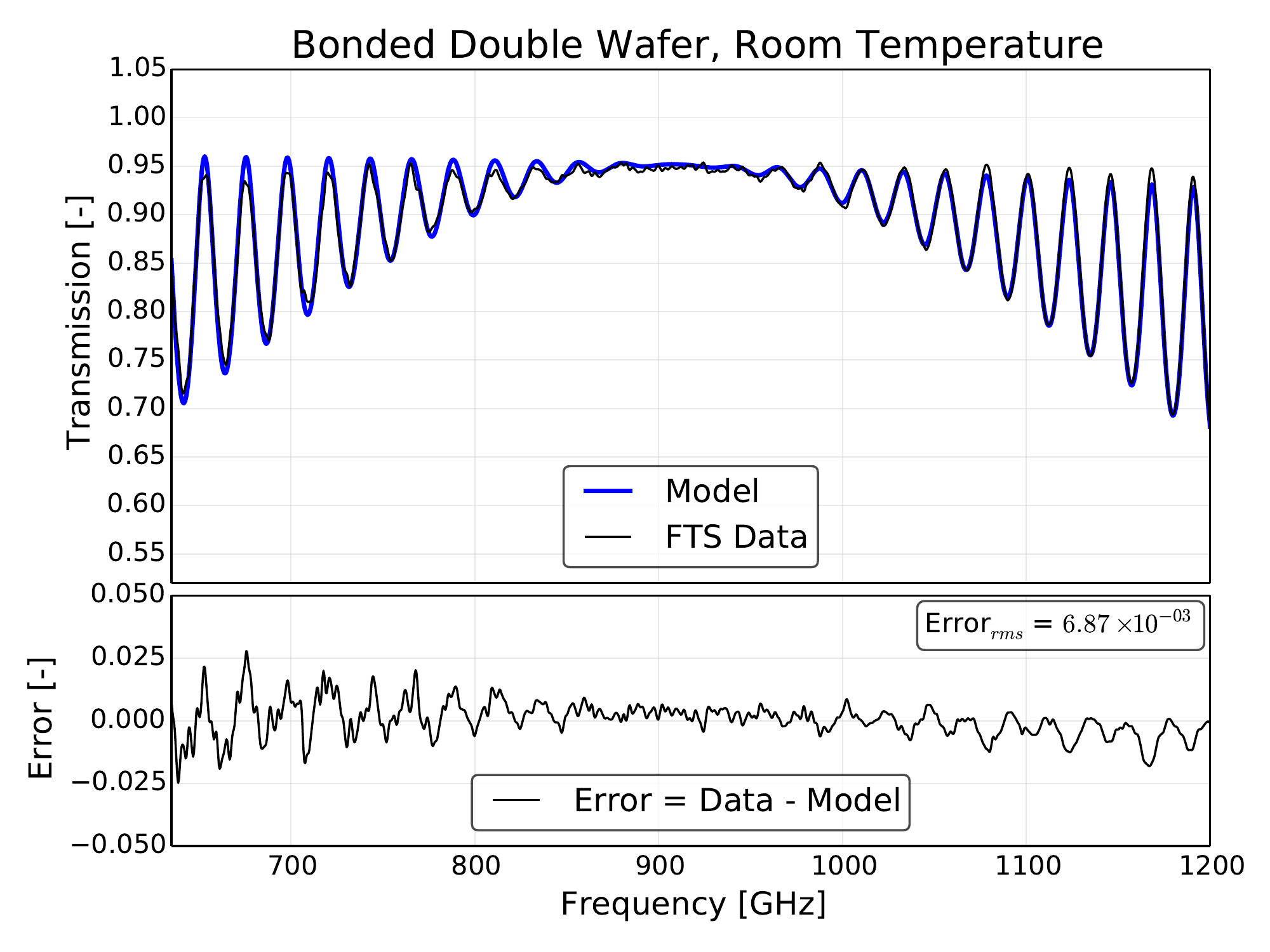}
\end{minipage}
\hspace{\fill}
\begin{minipage}{0.5\textwidth}
\includegraphics[width=\textwidth]{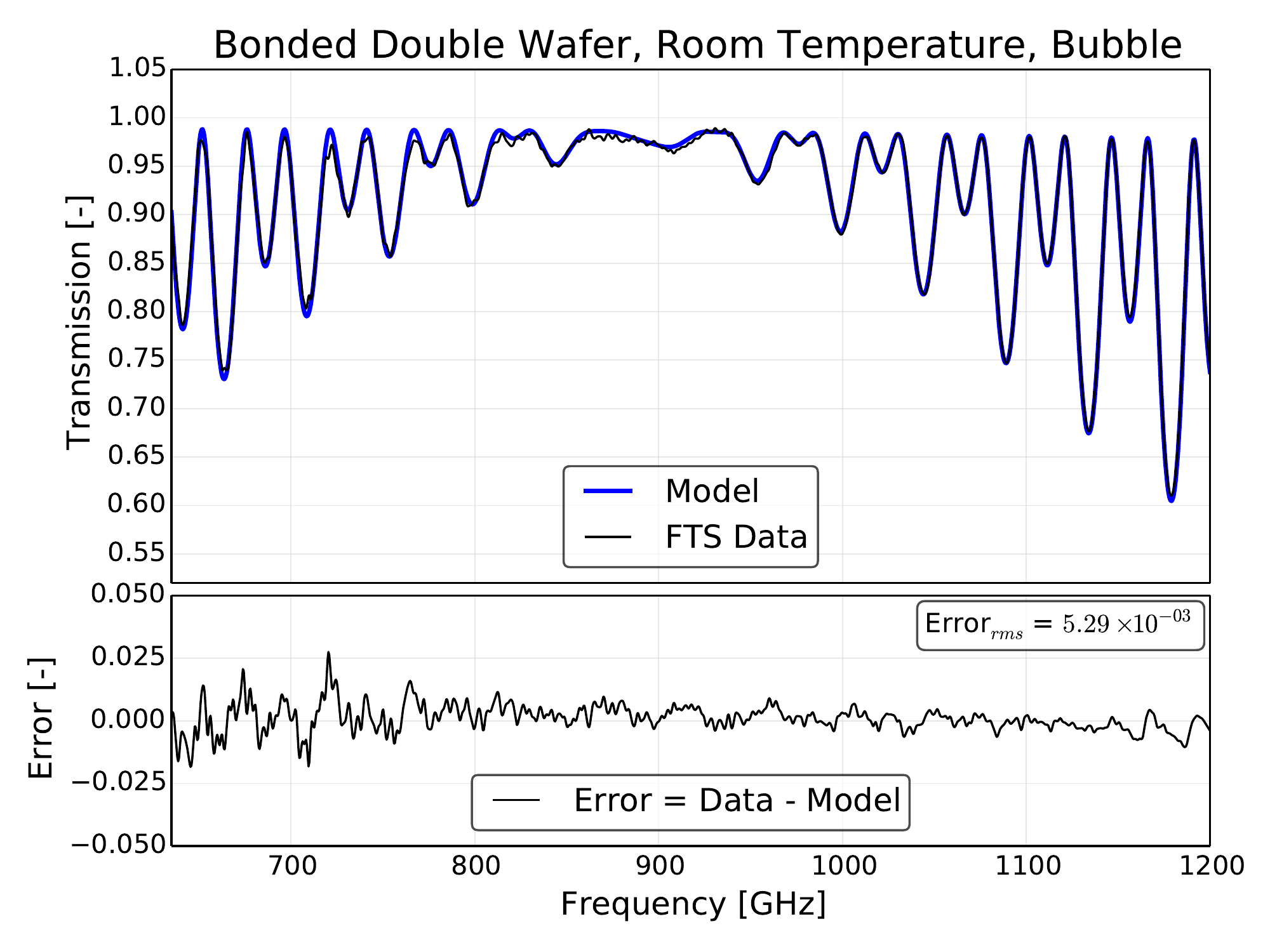}
\end{minipage}
\label{fig:Transmissions}

\caption{FTS transmission measurements on AR treated samples. Black line shows raw FTS data. Blue line shows our best fit parameter Fresnel model \cite{yeh_optical_1988}. Top Left: Room temperature sample coated on both sides (not bonded). Top Right: Cryogenic (4K) 2mm thick bonded sample. The high frequency ripples are due to reflections from the FTS cryostat sample chamber windows. Lower Left: Bonded silicon wafers at room temperature. Lower Right: Transmission measurements from portion of AR coated (and bonded) wafer with bubble defect.}  
\label{fig:Transmissions}
\end{figure*}

\subsection{Silicon Direct Bonding}

Bonded samples were made on 1 mm thick polished (on one side) wafers. These samples were bonded after they were AR coated with the DRIE process. Etching was performed on the rough side only and bonding was done on the polished sides.

Both etched wafers were put in an RCA clean \cite{kern1970cleaning} chemical treatment before bonding. This treatment consists of a basic solution ($H_2O:NH_4OH:H_2O_2 = 6:1:1$) and an acidic  ($H_2O:HCl:H_2O_2=6:1:1$) bath. Samples were put in each bath  for 10 minutes. They were centrifuged and dried right after the bath and put in contact immediately after drying while still in the clean room. 

Inside the furnace we used silica discs to press the wafers together. A silica cylinder was used to keep the wafers from sliding while in the furnace. We processed the samples in a high temperature (1100 C) furnace for two hours in a $N_2$ atmosphere. Ramp up and down times made  the entire furnace process take $\sim\!$ 1 day. After bonding, a Schott IR inspector at $\lambda = \SI{1.2}{\micro\metre} $ was used to look for bubbles between the bonded substrates. Figure \ref{fig:IR_Bonded} shows an IR image of the bonded sample. 

%% file: Sections/5fabresults.tex
\section{Process Characterization}

\begin{table*}[!htb]
\caption{\bf Best fit parameters obtained from FTS measurements shown in Figure \ref{fig:Transmissions}.}
\centering
\begin{tabular}{l | c | c | c | c | c}
Sample 			& $th_{AR} [\mu m]$ & $th_{Si}[\mu m]$ 	& $\epsilon_{AR}$ &$\epsilon''$ & $th_{gap}[\mu m]$ \\
\hline \hline
AR Warm & 42.93 & 437.0 & 3.547&   --  & --\\
AR Cold& 43.14&440.7 &3.526 &-- &--\\
Bonded Warm & 45.29 & 1928 & 3.288&$<4.2 \times 10^{-3}$ & --\\
Bonded Cold & 45.69 & 1928  & 3.30& $<3.3\times 10^{-4}$ & --\\
Bubble & 46.40 & 964.6* & $3.277$   & $<1.3 \times 10^{-3}$ &  2.71 \\
\end{tabular}
\label{tab:fitParams}
\\Parameters $\epsilon_{AR}$ and $\epsilon_{Si}$ are the real part of the dielectric constant of the AR layer and the bulk silicon respectively. $\epsilon''$ is the imaginary part of bulk silicon as defined in Equation \ref{eq:epsilon}. $th_{AR}$, $th_{Si}$ and $th_{gap}$ are the thicknesses of the AR layer, bulk silicon and the gap where the bonding did not yield. The bulk silicon dielectric constant was varied in the  allowed range in \cite{krupka_measurements_2006} and the other parameters were left free. Uncertainties from this fit are indicated in the last quoted digit. This approach neglects the metamaterial nature of the coatings, which leads to systematic uncertainties, such as the effective thickness variations discussed in \cite{Datta2013Large-apertureWavelengths}.
The silicon thickness fit in the five-layer bubble model is half the thickness in the three-layer model fit resulting in a consistent total thickness.
\end{table*}

The geometry of the etched samples was measured with a Zygo Optical profiler. The etched holes on a polished wafer can be characterized by a depth $d$  and radius of curvature $R$. The sharp corners in the square grid were smoothed out by the lithography process and can be described by a radius of curvature $r$ (see  Figure \ref{fig:structure} (top) for definitions). Measured values are shown in Table \ref{tab:metrology}. Unpolished samples show features on the etched holes. Figure \ref{fig:metrology} (left) shows a profile map of one of our measured samples which was fabricated on an unpolished wafer. Figure \ref{fig:metrology} (right) shows a series of histograms of profile maps taken over 1 square inch. The etched pattern does not deteriorate with the thermal process used during bonding or thermal cryogenic cycling.

Occasionally, the bonding process yielded bubbles where bonding was not possible. Figure \ref{fig:IR_Bonded} shows an IR image where one bubble was formed. These bubbles are believed to be generated by dust particles in the clean room and expanded in size due to the stiffness of silicon. A portable clean room setup can help improving bonding yield \cite{stengl_bubble-free_1988}.

%% file: Sections/6transmissionmeasurements.tex
\section{Transmission Measurements}

A Bruker IFS125 Fourier Transform Spectrometer (FTS) was used. The transmission geometry configuration was selected to test the performance of the AR layer and the performance of the bonded AR treated sample over the frequency range 0.9 to 21 THz. The FTS employed a 6 micron mylar beamsplitter in conjunction with a mercury arc lamp and a liquid helium cooled bolometer detector to span the aforementioned frequency band. Samples were mounted in an Oxford liquid helium cooled cryostat for measurements at sample temperatures of 10 K. The single silicon piece etched on both sides was used to test the transmission properties of the AR coating alone. The bonded wafers were tested to study the transmission properties of the entire process (DRIE AR treatment and high temperature bonding).

\subsection{Properties of silicon}
Transmission was modeled using a multiple layer model for effective media \cite{yeh_optical_1988, choy_effective_2015}. Here, each layer has a (effective) dielectric constant and a finite thickness. The dielectric constant $\epsilon$ is related to the index of refraction $n$ according to 
\begin{equation}
n^2 = \epsilon = \epsilon' + i \epsilon''
\label{eq:epsilon}
\end{equation}
The transmission line matrix formulation for isotropic layered media presented in \cite{yeh_optical_1988} was used to describe the transmission coefficient $T$ of a stack of layers. Where $T=1/M_{11}$ for normal incidence and $M$ is the matrix that represents the stack of layers in this formalism.

In particular, a three layer (AR-Si-AR) model was used. In this model each layer has a constant dielectric function $\epsilon$. The dielectric constant of the bulk silicon is taken from \cite{krupka_measurements_2006} while the dielectric constant of the AR layer and its thickness are fitted from the FTS transmission data.

\subsection{Single wafer tests}

FTS measurements on the double sided AR treated single silicon piece show excellent  transmission (better than $99\%$) both at room temperature and at cryogenic temperatures (10 K) which is consistent with the high resistivity specs of this sample ($>2000\ \Omega\ \mathrm{cm}$). 

The best fit model parameters are shown in Table \ref{tab:fitParams}. Figure \ref{fig:Transmissions} (upper left) shows the transmission measurements. Transmission better than $99\%$ is achieved between 840 and 980 GHz. Fit residuals, have an RMS error of $0.7\%$.

\subsection{Bonded wafer tests}

Cryogenic measurements show a transmission better than $99\%$ from 876 to 934 GHz. Figure \ref{fig:Transmissions} (upper right) shows the best fit model (Table \ref{tab:fitParams}) and the FTS transmission data. High frequency ripples are an artifact of the measurement and resulted from the cryogenic window used. RMS error is $1.4 \%$ from 600 to 1250 GHz.

Room temperature measurements show a loss of approximately $5\%$ as shown in Figure \ref{fig:Transmissions} (lower left). This room-temperature loss was not expected for the roughly $1000\ \Omega\ \mathrm{cm}$ resistivity silicon used for these samples and is under investigation. It is clear that the loss becomes negligible (see Figure \ref{fig:Transmissions} (upper right)) at cryogenic temperatures (10 K) where similar optics often operate.

\subsubsection{Effect of gaps between bonded surfaces}

Transmission in the sample that shows the largest $\sim 1 \ \mathrm{cm} $ bubble as seen in the IR image (Figure \ref{fig:IR_Bonded}) is measured. The transmission obtained shows a double peak pattern (Figure \ref{fig:Transmissions}, lower right). This fringe pattern can be modeled with a 5 layer model (AR-Si-Air-Si-AR). Our best fit parameters (Table \ref{tab:fitParams}) show that this model is consistent with a $\SI{2.7}{\micro\metre}$ air gap between the wafers. Fringe counts seen in the IR image suggest an upper bound of $\sim \SI{4}{\micro\metre}$ for this thickness, consistent with direct measurements of the thickness of the stack on this region and with the average gap size illuminated in the FTS measurements.

%% file: Sections/7twolayer.tex
\section{Prototype Two-layer Coating}

\label{sec:pillars}

\begin{figure}
\centering
    \includegraphics[width=\linewidth]{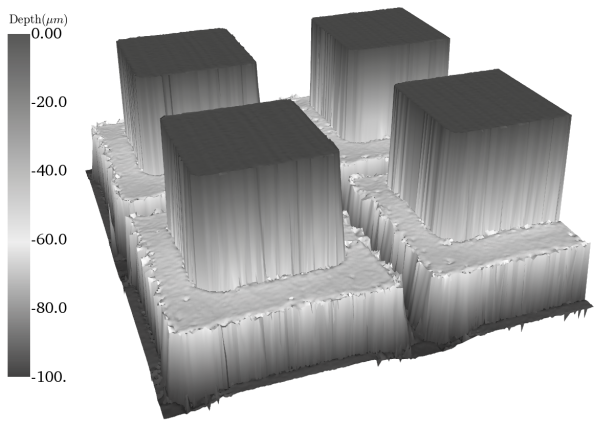}
  \caption{Prototype two-layer coating with a pillar geometry scaled from \cite{Datta2013Large-apertureWavelengths} fabricated using DRIE. The 3D image is reconstructed based on depth measurements with a Zygo optical profiler.}
  \label{fig:2layer}
\end{figure}

Many silicon optics applications stand to benefit from wider bandwidth and reduced reflections than are provided by a single-layer AR coating. A prototype two-layer coating comprised of pillars has been developed by scaling the coating design in \cite{Datta2013Large-apertureWavelengths} for a wavelength of \SI{350}{\micro \metre} (Figure \ref{fig:2layer}). 

The prototype two-layer AR coating was fabricated on a silicon wafer by using two etch masks to define two sets of trenches of different widths and depths. The first etch mask was created by depositing \SI{500}{\nano \metre} of silicon oxide onto the wafer using plasma enhanced chemical vapor deposition (PECVD) and etching it into small squares using Reactive Ion Etching (RIE). These small squares of oxide define the top of the upper pillar. The second etch mask was created using photolithography to pattern larger squares on top of the oxide squares. These larger squares define the top of the lower pillar. The wafer was then processed using DRIE to etch the lower trenches to a depth of \SI{42}{\micro \metre}. The photoresist was removed, exposing the silicon oxide, and the wafer was etched again until the upper trenches reached a depth of \SI{62}{\micro\metre}. During this step, the lower trenches were etched to a depth of \SI{31}{\micro \metre} below the upper trenches. The lower trench depth was effectively raised by \SI{11}{\micro\metre} due to a mismatch in etch rates between the two trenches. The oxide etch mask was removed in an HF bath. 

A next step in developing multi-layer coatings like this could include more precise characterization of the relative etch rates for the lower and upper trenches to optimize the AR coating performance. This approach shows significant promise for fabricating broader bandwidth coatings in the future.

%% file: Sections/8conclusion.tex
\section{Conclusion}
We have developed single-layer silicon AR coatings at sub-millimeter wavelengths and a prototype two-layer coating using a DRIE technique on 100 mm diameter silicon wafers. A Silicon Direct Bonding approach was used to bond the AR coated silicon to a substrate for use in refractive optical elements. No glue is used in this process. We have shown that this method does not introduce additional losses at the  percent-level. A few micron bubble was formed during the bonding process and characterized, with suggestions made for eliminating bubble formation in the future.

Optical elements that could utilize this technique directly include Fabry-P\'{e}rot cavities, half-wave plates, and the planar side of plano-convex lenses. 

Extensions of this technique include full development of multi-layer AR coatings for sub-millimeter and millimeter wavelengths and implementation on larger diameter wafers. Similar bonding techniques could also be used to bond AR-coated wafers to the curved surfaces of lenses \cite{aono_large_2013}.